\documentclass[a4paper]{jpconf}
\usepackage{graphicx}
\begin{document}
\title{Continuing Progress on a Lattice QCD Software Infrastructure}

\author{B\'alint Jo\'o on behalf of the USQCD Collaboration}

\address{Thomas Jefferson National Laboratory, 12000 Jefferson Avenue, Newport News, VA 23606, USA}

\ead{bjoo@jlab.org}

\begin{abstract}
We report on the progress of the software effort in the QCD Application
Area of SciDAC. In particular, we discuss how the software developed
under SciDAC enabled the aggressive exploitation of leadership
computers, and we report on progress in the area of QCD software for multi-core
architectures.
\end{abstract}

\section{Introduction}
Large scale numerical lattice QCD simulation programs require
extensive software infrastructure \cite{Joo}. 
 In this contribution
we report on some advances in the
 SciDAC supported software work
within the USQCD national program \cite{USQCD} during the past year. 
Details about the scientific results from numerical QCD simulations
can be
 found in \cite{Kronfeld}.
 
 Considerable effort was
invested this past year in preparation
 for the exploitation of
leadership computing facility at Argonne (ANL) and Oak Ridge (ORNL)
National Laboratories. We have also
 carried out investigations into
the efficient use of multi
 core architectures for QCD. Our data
sharing efforts have progressed,
 and we continue to extend and
refine standards within our
 collaboration and worldwide.  
 This
article is organized as follows: in section \ref{s:LCF} we
 present
some highlights of our exploitation of leadership computing
resources. We report on our threading research in section
\ref{s:Threading}. We discuss data sharing in section
\ref{s:DataSharing} and briefly consider other activities in section
\ref{s:Miscellany}.
 
\section{Exploiting Leadership Facilities for QCD}\label{s:LCF}
During the last year, all of our large community codes have been
successfully ported to leadership systems of interest. In particular,
the Chroma \cite{Chroma} and MIMD Lattice Collaboration \cite{MILC} (MILC) 
codes have been
 deployed on Cray XT, BlueGene/P and recent cluster hardware
while the Columbia Physics System (CPS) \cite{CPS} has been ported to BlueGene/P.
Considerable effort has been invested in the optimization of high
performance components for these architectures.

\begin{figure}[ht]
\begin{center}
\includegraphics[height=2.4in]{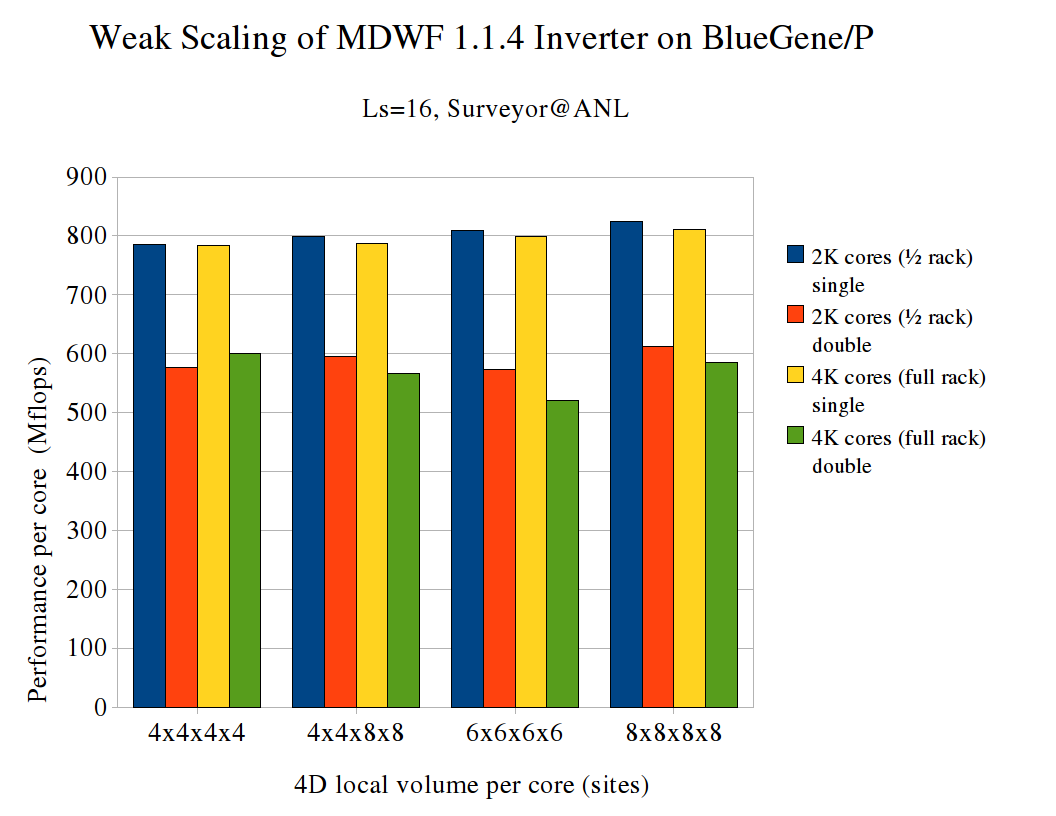}
\includegraphics[height=2.4in]{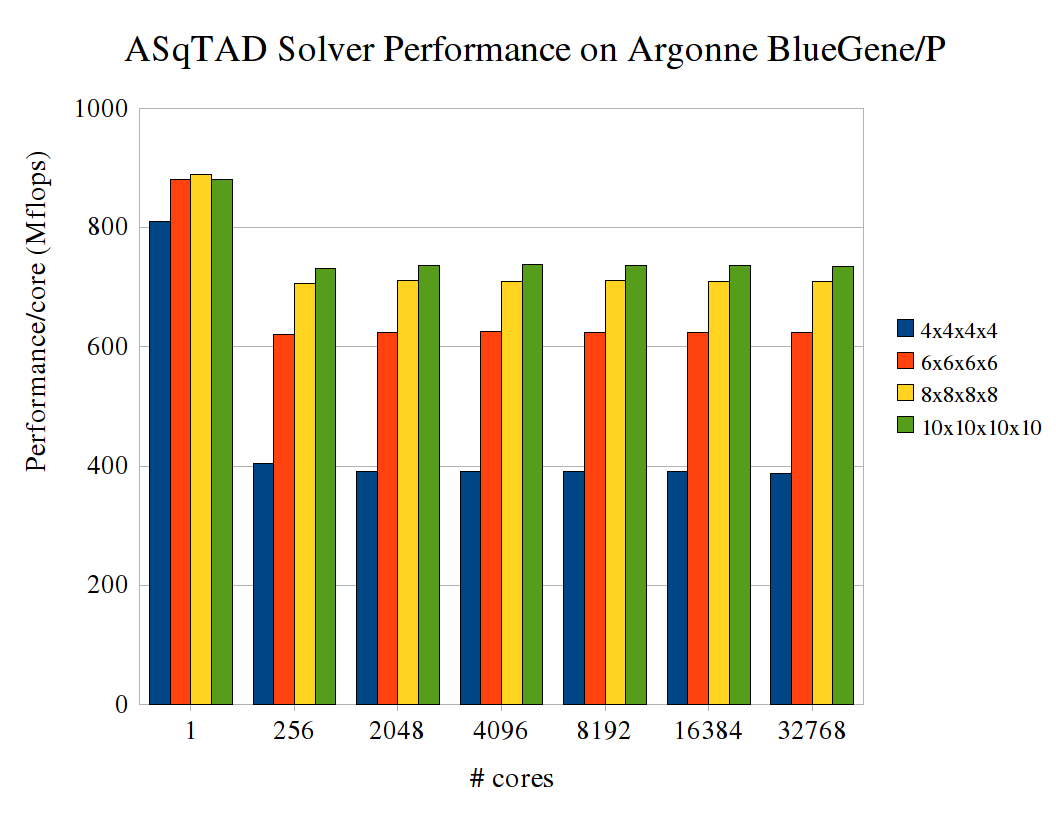}
\end{center}
\vspace{-0.5cm}
\caption{\label{f:Perf}Performance on BlueGene/P for MDWF \cite{MDWF} (left) and AsqTAD (right) inverters. The MDWF measurements were made with Chroma interfaced to MDWF 1.1.4 on Surveyor. The AsqTAD numbers are from the MILC Code on Intrepid, both machines being at ALCF ANL}
\end{figure}

We show in the left half of figure \ref{f:Perf} the performance of the
MDWF solver on
 the BlueGene/P at ANL.  MDWF is an optimized Domain
Wall Fermion (DWF) Conjugate Gradients
 Inverter package -- developed
at MIT \cite{MDWF} -- which incorporates several DWF Operator
variants. One can see that the single precision solver achieves some
800 Mflops/core (about 25\% of peak). The solver performance is
roughly constant over a variety of volumes, indicating good
strong-scaling and that the recursive data ordering on node results in
effective use of cache for the larger problems. MDWF has been interfaced with 
Chroma for production use, by Jefferson Lab (JLab) staff.

The MILC Collaboration have optimized their AsqTAD Conjugate
Gradients
 Solver for the BlueGene/P and the Cray XT series, and
their gauge force for a variety of platforms. We show the performance
of their AsqTAD inverter on the Argonne BlueGene/P in the right hand
plot of figure \ref{f:Perf}. One can see that the weak scaling is
excellent all the way out to 32K cores.


\begin{figure}[ht]
\begin{center}
\includegraphics[height=2.6in]{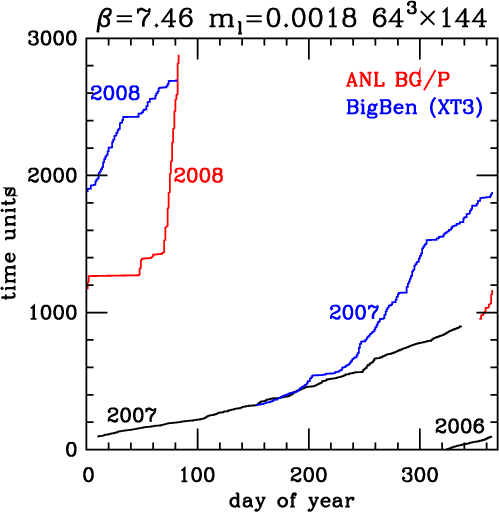}
\includegraphics[height=2.6in]{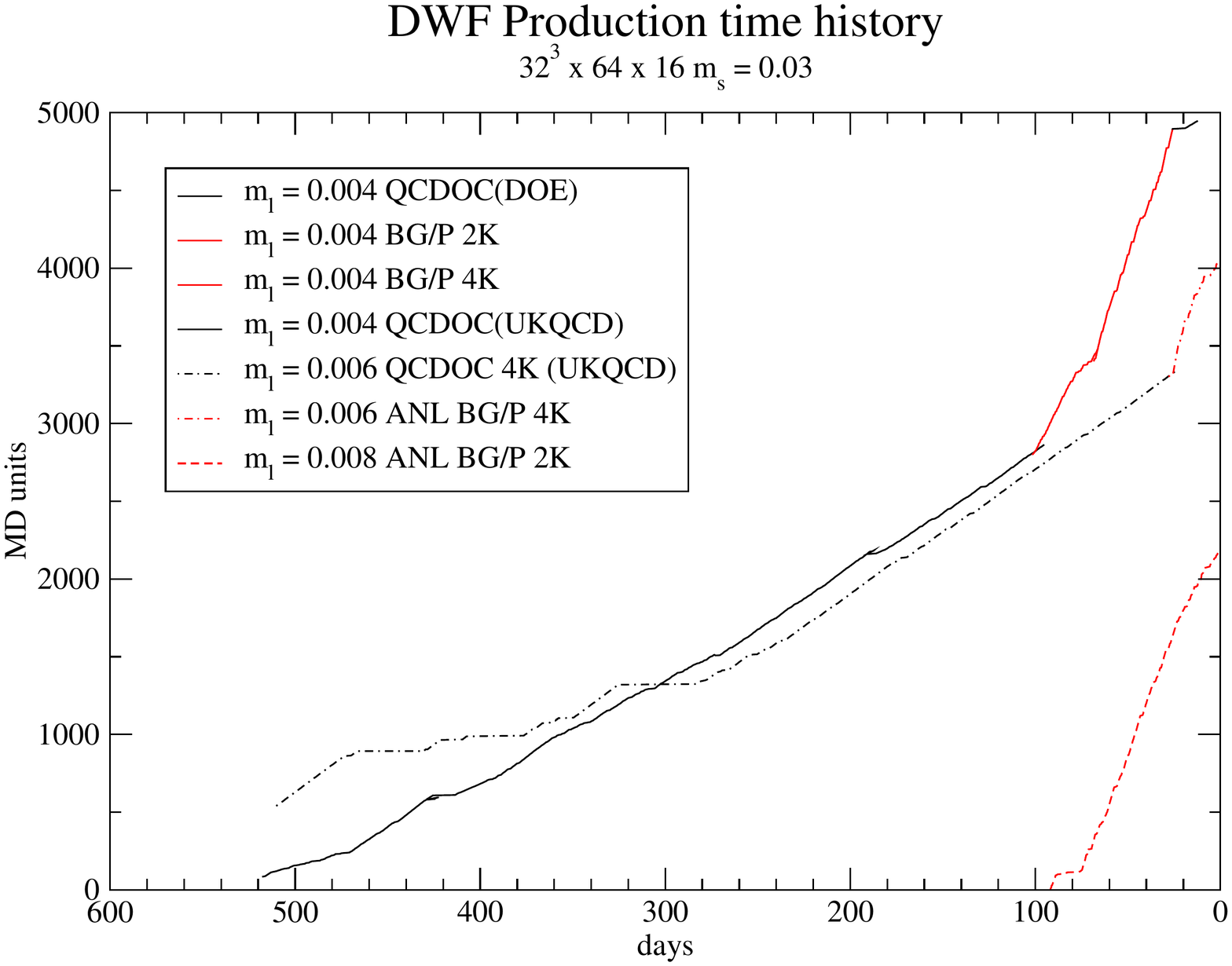}
\end{center}
\vspace{-0.5cm}
\caption{\label{f:Hist}Production history of an AsqTAD Staggered Fermion run through late 2006, 2007 and early 2008(left) and for some Domain Wall Fermion production runs (right).}
\end{figure}
 
To emphasize the impact the leadership machines have had on our data
production we show in figure \ref{f:Hist} the production history of 
both an AsqTAD and some DWF runs over the past year.
The AsqTAD production was performed using the MILC code while the
Domain Wall Fermion production was carried out by colleagues based 
at Columbia University and Brookhaven National Laboratory (BNL) using the 
CPS code. The dramatic increases in production brought about by the BlueGene/P 
for both sets of data production (red lines) are clearly evident. 


In summary, USQCD has completed successful ports of its applications to
leadership hardware, and is now reaping the benefits in terms of
science production.

\section{Threading Investigations}\label{s:Threading}
\begin{figure}[ht]
\begin{center}
\includegraphics[width=3.5in]{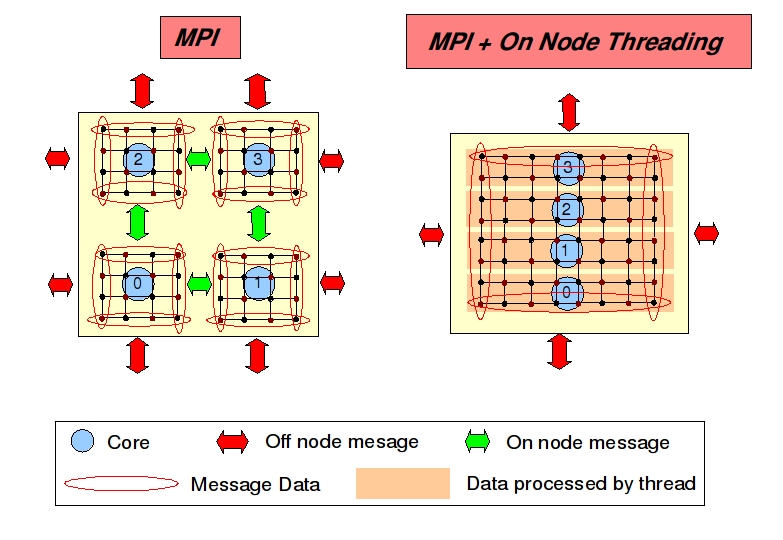}
\end{center}
\vspace{-0.5cm}
\caption{\label{f:ThreadSavings}Communications patterns of the Wilson Dslash kernel in 2D on a quad core node for a full MPI application (left) and a Hybrid Threaded-MPI application (right)}
\end{figure}

In preparation for the arrival of multi-core hardware, JLab staff,  
in collaboration with EPCC in Edinburgh UK, have added
threading support to one of our key computational kernels, called {\em
  Wilson Dslash} \cite{Dslash}. This kernel is a 4 dimensional
nearest neighbor operator. We show a 2D schematic picture of the
communications patterns in this operator in figure
\ref{f:ThreadSavings} for a pure MPI implementation on the left,
versus a Hybrid-threaded one on the right. Two potential efficiency
gains are immediate: First, one can see that threading on-node
eliminates on-node messaging (green arrows in the figure). Second, 
the threading effectively coalesces multiple messages
that would have been sent by individual cores into fewer, larger messages
sent by the node (red arrows, ellipses in figure) which may be
advantageous in some networks.

Multi-threading on node is typically realized using OpenMP or some custom
thread library. We have developed a lightweight thread library called
QCD Multi--Thread (QMT) \cite{QMT} which enables a data parallel
programming technique similar in spirit to OpenMP: Work is supplied to QMT 
by calling the {\tt qmt\_call()} library function, with a
call back procedure that can perform part of the desired work. QMT then
invokes this function, with different parts of the problem from
different threads. When {\tt qmt\_call()} completes, it calls a
barrier among the threads to synchronize them.  
In the work described here we used QMT with a
queue based barrier, optimized for the MOESI cache coherence protocol
on AMD Barcelona cores.

Our numerical experiments consisted of running the Wilson Dslash
operator on several nodes of the Jaguar Cray XT4 system at ORNL,
either as a pure MPI or as a hybrid MPI-threaded application, using
alternately both OpenMP and QMT for the threading. We performed
our tests on a single node and then repeated them on 16 nodes which
could be mapped as a $2^4$ processor grid
communicating in 4 directions. The tests were repeated using several
local volumes: $2^4$, $4^4$, $6^4$ and $8^4$ respectively. In
particular the $2^4$ volume is our hard scaling limit with all sites on the
surface and the $8^4$ volume is typically too large to be cache resident. We have
found the $6^4$ volume most efficient with $4^4$ volume case less so
due to a worse surface to volume ratio.

\begin{figure}[ht]
\begin{center}
\leavevmode
\hbox{
\hspace{-1.3cm}
\begin{tabular}{c}
 \includegraphics[height=2.8in]{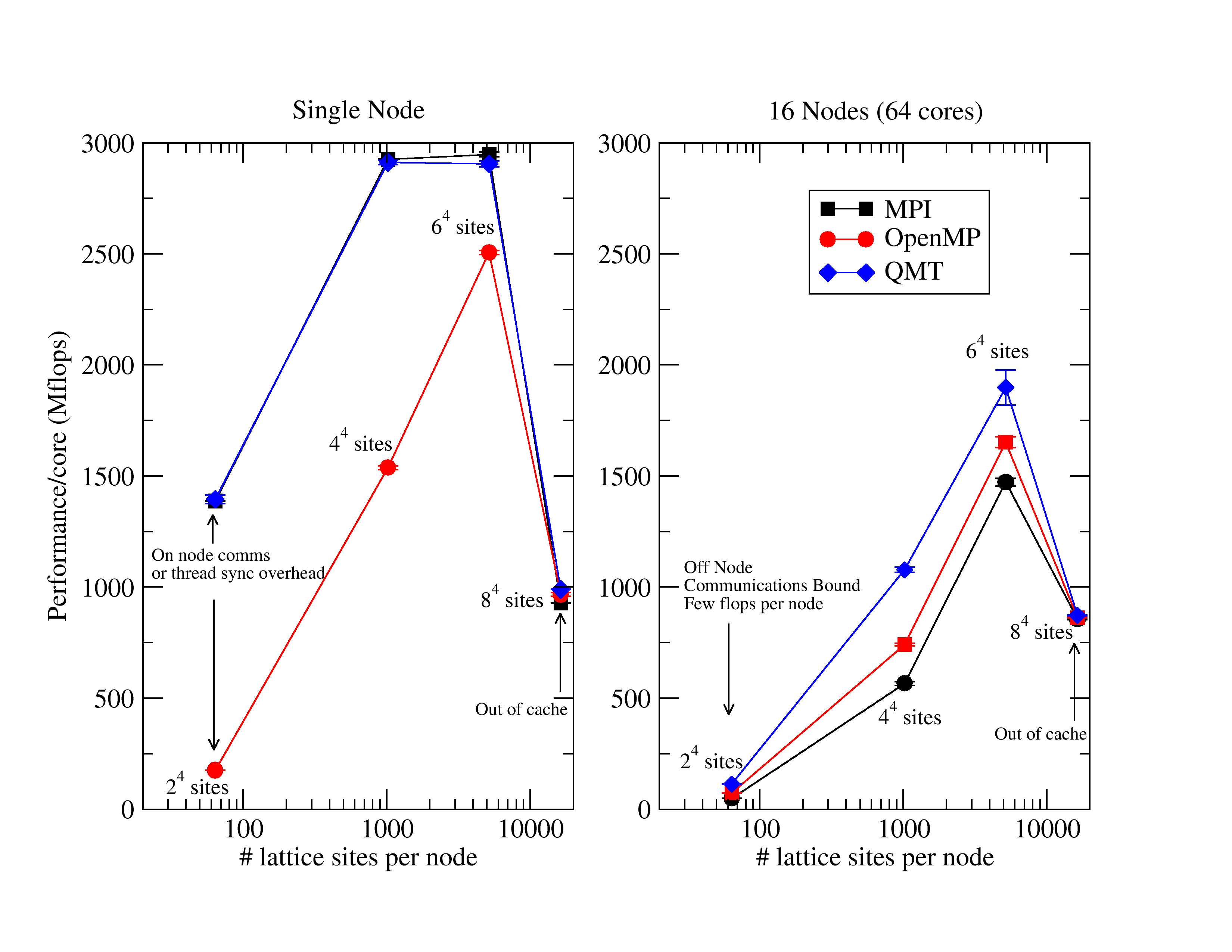} \vspace{-0.5cm}\\
  a) 
\end{tabular}
\hspace{-1.5cm}
\begin{tabular}{c}
 \includegraphics[height=2.8in]{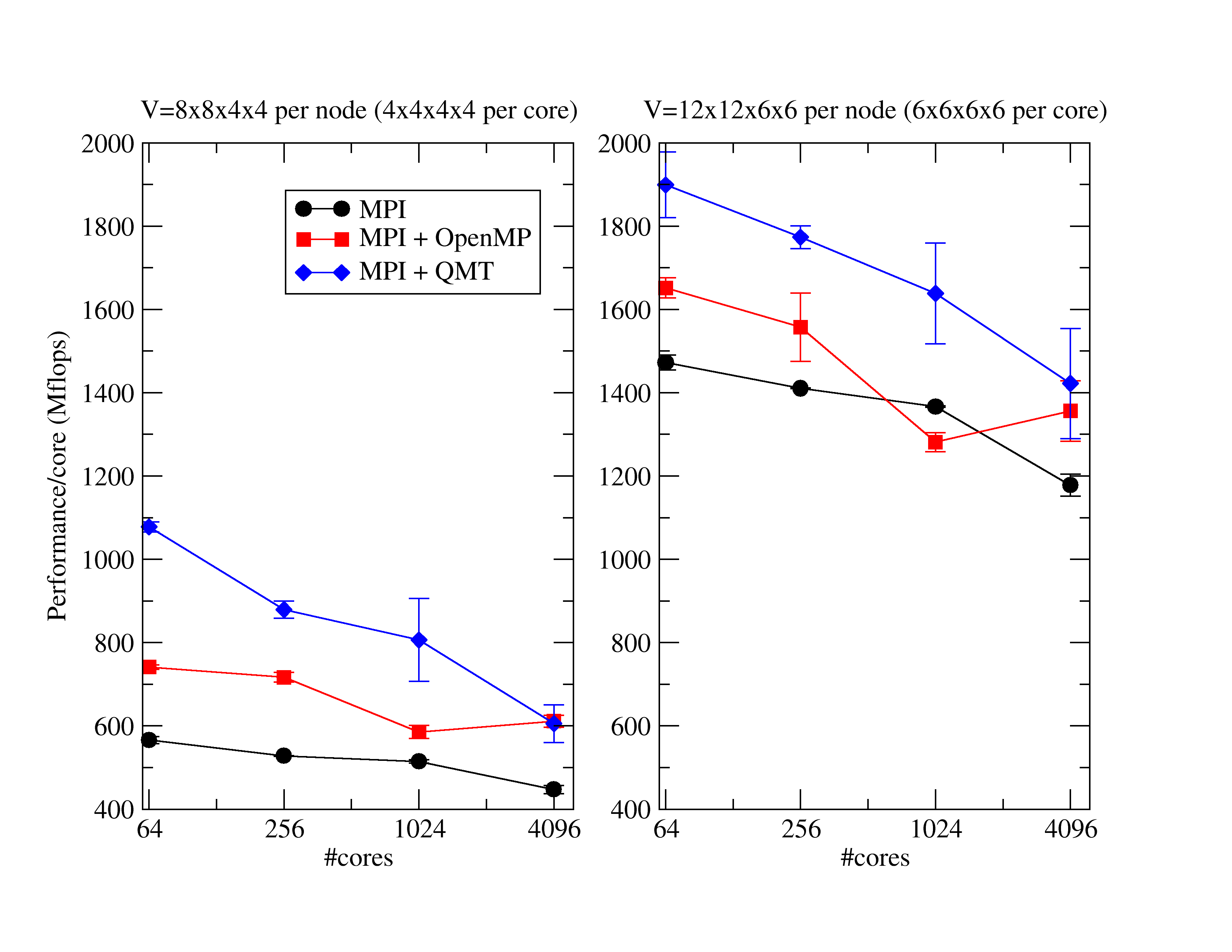} \vspace{-0.5cm} \\
 b)
\end{tabular}}
\end{center}
\vspace{-0.5cm}
\caption{\label{f:Threading}a)The effects of threading the Wilson Dslash operator on a single node (left) and a partition of 16 nodes (right) of Jaguar. b)Weak scaling of the performance of the Wilson Dslash operator on Jaguar for two fixed local volumes: $4^4$ sites (left) and $6^4$ sites (right), to 4096 cores.}
\end{figure}

Our results are shown in figure \ref{f:Threading}. It can be seen in figure \ref{f:Threading}a) that on a single node
the QMT and MPI performances are essentially identical. 
Looking at results for 16 nodes, we see that the threaded
performances are much improved for the $4^4$ and $6^4$ volumes over
the pure MPI case. In the cases of the smallest and largest volumes
there seems to be no difference between the pure MPI and hybrid
threaded versions. In the situations where threading results in a gain, 
using QMT for on-node threading results in a higher performance than 
when using OpenMP.

We then performed a weak scaling benchmark for the $4^4$ and $6^4$
local volumes in an attempt to scale the gains from the Hybrid-Threaded approach 
up to a large partition. Our results are shown in figure \ref{f:Threading}b). One can see that for
both local volumes, a performance advantage is maintained over pure
MPI for as far out as 4096 cores when using QMT. When
using OpenMP, the weak scaling appears quite erratic, but typically
performance is less than the QMT case except for the the 4096 core
partition size.


Our interpretation of these results is as follows: Since single
node tests suggest no gain from eliminating on-node messages, we
surmise that in our application,
 threading gains performance over
pure MPI due to the collation of off-node messages.  This gain is
likely to be network and OpenMP implementation dependent.  Our
limiting $2^4$ volume is completely communications bound with very few
flops to overlap with communication, hence the low performance in that
case for all the approaches tried. In the cases of the $4^4$ and $6^4$
volumes,
 message collation reduces the number of messages and
increases
 their size, thus taking better advantage of the Cray
network. In the 
 $8^4$ volume case we fall out of cache and all
approaches appear to perform
 equally.


 In this investigation, we have neglected issues that arise in
multi-socket NUMA architectures such as thread, process and memory
affinity. We have also not explored general multi--core aspects such
as
 the abundance of floating point power versus the comparative 
lack of memory bandwidth.  Partitioned Global Address Space languages
such as UPC provide a natural programming model for NUMA based
architectures, and aggressive prefetching and double
 buffering may
alleviate the memory bandwidth issue to some degree.  Some of these
concerns are investigated in \cite{Fowler} and we intend to explore
these issues more fully in future work.

\section{Data Sharing and Grid Related Efforts} \label{s:DataSharing}

 Our data sharing efforts have continued in the past year by
publishing
 many of our gauge configurations on the International
Lattice Data
 Grid (ILDG)\cite{ILDG}, and some through other channels
\cite{BNLLat,GaugeConn}. Our ILDG infrastructure is based jointly at Fermi
National Accelerator Laboratory (FNAL) where the storage element is
hosted and managed, and at JLab where one can find the Metadata and
File Catalog Web Services.  Some 11,000
configurations are now published in 16 ensembles through the ILDG. Conversely,
in the past year several
 US researchers have joined the ILDG virtual
organization, in order to use published data shared through ILDG.
 
 Within USQCD data sharing has moved forward
through the definition of
 file formats for quark propagators and the
implementation of software
 to read and write the standard within the
QIO library. Application
 codes have also been modified to read and
write these files. European
 collaborators have defined propagator
formats that are compatible with
 the USQCD format and there is some
hope that worldwide propagator
 sharing will eventually be formalized
in the ILDG.

\section{Other Activities} \label{s:Miscellany}

The USQCD software program, continues its work and collaboration in many other areas not discussed here for lack of space, including the application of Workflows to QCD, algorithmic developments, improvements to data analysis, code optimization, visualization and the use of Grid technologies. 

\section{Summary}

In this contribution, we presented an overview of software progress in
lattice QCD in the US over the past year, with emphasis on performance
achieved on leadership computers, and our work on multi--core
architectures.  Much of this work was carried out in international
collaboration, in particular with colleagues in the UK.
We intend to continue progress in the software
area in the future in order to carry on with our highly successful
exploitation of available resources and to
continue producing high quality scientific results and discoveries from lattice QCD.

\section{Acknowledgements}
The author thanks Steve Gottlieb (Indiana University), James Osborn (ALCF, ANL) and Chulwoo Jung (BNL) for providing the data and plots for figures \ref{f:Perf} and \ref{f:Hist}. The MDWF Inverter was written
by Andrew Pochinsky at MIT and can be downloaded from \cite{MDWF}.

This research used resources of the National Center for Computational Sciences at Oak Ridge National Laboratory, which is supported by the Office of Science of the Department of Energy under Contract DE-AC05-00OR22725. This research used 
resources of the Argonne Leadership Computing Facility at Argonne National Laboratory, which is supported by the Office of Science of the U.S. Department of Energy under contract DE-AC02-06CH11357. 

The USQCD Software effort is funded by the SciDAC program of the U.S. Department of Energy.


\end{document}